\documentclass[preprint]{elsarticle}

\usepackage{graphicx}
\usepackage{amssymb}
\usepackage{tabularx}
\usepackage{xcolor}
\journal{Journal of Magnetism and Magnetic Materials}

\begin{document}

\begin{frontmatter}

\title{Structural, electronic and magnetic properties of La$_{1.5}$Ca$_{0.5}$(Co$_{0.5}$Fe$_{0.5}$)IrO$_{6}$ double perovskite}

\author[AA]{L. Bufai\c{c}al\corref{Bufaical}}, \ead{lbufaical@ufg.br}, \author[BB]{M. A. V. Heringer}, \author[BB]{J. R. Jesus}, \author[BB]{A. Caytuero}, \author[BB]{C. Macchiutti}, \author[BB]{E. M. Bittar}, \author[BB]{E. Baggio-Saitovitch}

\address[AA]{Instituto de F\'{i}sica, Universidade Federal de Goi\'{a}s, 74001-970 , Goi\^{a}nia, GO, Brazil}

\address[BB]{Centro Brasileiro de Pesquisas F\'{\i}sicas, Rua Dr. Xavier Sigaud 150, 22290-180, Rio de Janeiro, RJ, Brazil}

\cortext[Bufaical]{Corresponding author}

\begin{abstract}
In this work, we report the synthesis and investigation of structural, electronic, and magnetic properties of La$_{1.5}$Ca$_{0.5}$(Co$_{0.5}$Fe$_{0.5}$)IrO$_{6}$. Our polycrystalline sample forms as a single-phase double perovskite in monoclinic $P2_{1}/n$ space group. Co and Ir are most likely in bivalent and tetravalent oxidation states, respectively, while M\"{o}ssbauer spectroscopy indicates that Fe is in a trivalent state. The ac and dc magnetization data suggest a ferrimagnetic behavior resulting from the presence of two antiferromagnetic sublattices at Co/Fe and Ir sites. The large coercive field $H_C$ $\simeq$ 32 kOe observed at 10 K, comparable to that of other double perovskites of interest for hard magnets, is discussed in terms of the structural distortion and the spin and orbital magnetic moments of the transition metal ions.
\end{abstract}


\end{frontmatter}

\section{INTRODUCTION}

In the $A_2BB’$O$_6$ double perovskites (DP), the possibility of accommodation of several distinct elements in both A site with alkaline/rare-earth ions and in $B$ and $B’$ sites with transition metal (TM) ions makes this one of the most extensively investigated structures in the last decades \cite{Sami}. Different combinations of $A$ and $B/B’$ ions can lead to interesting physical properties such as, for instance, multiferroicity \cite{Azuma}, exchange bias \cite{APL2020} and even possibly superconductivity \cite{Dow}. 

With respect to magnetic properties, the most interesting phenomena are usually observed for combination of 3$d$ with 4$d$/5$d$ TM ions at $B$ and $B’$ sites, leading for instance to high temperature ($T$) ferrimagnetism in Sr$_2$CrOsO$_6$ \cite{Krockenberger}, half-metallic behavior in Sr$_2$FeMoO$_6$ \cite{Tokura} and giant magnetoresistance in Mn$_2$FeReO$_6$ \cite{Li}. Despite that, the use of 5$d$ Ir ion in DP compounds have received less attention for many years. But recently, the discussion concerning the existence of excitonic magnetism in the anticipated nonmagnetic $j$ = 0 ground state of 5$d^4$ Ir$^{5+}$ ions in Sr$_{2}$YIrO$_{6}$6 and Ba$_{2}$YIrO$_{6}$ have put Ir-based systems in the spotlight \cite{Cao,Corredor,Fuchs}. 

There are, however, some earlier works reporting interesting physical properties due to the extended Ir 5$d$ orbitals. For instance, electronic structure calculations have predicted a metallic ground state in La$_2$CoIrO$_6$. But this was not confirmed by experimental results, which showed a FM-like insulating state for this Co- and Ir-based DP \cite{Narayanan,PRB2020}. Further studies of this material revealed magnetodielectric effect together with re-entrant spin-glass behavior at low-$T$ \cite{Song}, whereas Ca$^{2+}$ to La$^{3+}$ partial substitution leads to compensation temperatures and spontaneous exchange bias effect in La$_{1.5}$Ca$_{0.5}$CoIrO$_6$ (LCCIO) \cite{PRB2016}. For the case of Fe- and Ir-based DPs, the Ca$^{2+}$ to La$^{3+}$ partial substitution in La$_{2-x}$Ca$_x$FeIrO$_6$ induces interesting changes in the nature of the microscopic magnetic interactions between the TM ions, where the system evolves from antiferromagnetic (AFM) in the $x$ = 0 and 2.0 extremities of the series to FM-like in the $x$ $\sim$ 1.0 intermediate region, this being ascribed to changes in Ir formal valence \cite{JAP2008}.

The delicate balance between the strong spin-orbit coupling (SOC), the Coulomb repulsion, and the crystal field splitting in Ir at octahedral coordination makes the Ir-based DPs very sensitive not only to hole/electron doping at $A$-site, but also to situations where the chemical doping with ions of same oxidation state acts mainly to change the crystal structure. For example, Lu$_2$NiIrO$_6$ is a ferrimagnetic (FIM) Mott insulator where the Lu smallest ionic radius among the $A_2$NiIrO$_6$ ($A$ = La, Pr, Nd, Sm, Eu, Gd, Lu) family leads to the largest structural distortion that in turn enhance the AFM coupling between half-filled Ni-$e_g$ and partially-filled Ir-$t_{2g}$ orbitals, resulting in its higher $T_C$ \cite{Feng}. 

Doping at the $B$ sites in Ir-based DPs may also lead to interesting results. For instance, very recent density functional theory (DFT) calculation in the aforementioned Lu$_2$NiIrO$_6$ compound has predicted that 50\% doping with Cr, Mn or Fe at Ni site results in an electronic transition from Mott-insulating to half-metallic state, in which the admixture of Ir 5$d$ orbitals in the spin-majority channel are mainly responsible for the conductivity, while the spin minority channel remains an insulator \cite{Nazir}.

In this work, we investigate the combined effect of doping at both $A$- and $B$-sites in an Ir-based DP. The structural, electronic and magnetic properties of La$_{1.5}$Ca$_{0.5}$(Co$_{0.5}$Fe$_{0.5}$)IrO$_{6}$ (LCCFIO) polycrystalline sample were studied by means of X-ray powder diffraction (XRD), ac and dc magnetization measurements and M\"{o}ssbauer spectroscopy. Our results show that LCCFIO forms as a single-phase DP in monoclinic $P2_{1}/n$ space group. M\"{o}ssbauer spectroscopy indicates that Fe is in the trivalent state, while magnetization as a function of $T$ and applied field ($H$) curves suggest mixed-valence for Co and Ir. The magnetometry also revealed an FM-like weak magnetization below 112 K, possibly an FIM behavior, which is mainly discussed in terms of a noncollinear magnetic (NCM) structure where Co/Fe and Ir form two AFM sublattices. The large coercive field $H_C$ $\simeq$ 32 kOe observed at low-$T$ is discussed in terms of the structural distortion and of the valence states of TM-ions.

\section{EXPERIMENT DETAILS}

A polycrystalline sample of LCCFIO was synthesized by the conventional solid-state reaction method. Stoichiometric amounts of La$_{2}$O$_{3}$, CaO, Co$_{3}$O$_{4}$, Fe$_{2}$O$_{3}$ and metallic Ir in powder form were mixed and heated at $800^{\circ}$C for 12 hours in air atmosphere. Later the powder was mixed before a second step at $1200^{\circ}$C for 24 hours. Finally, the material was ground, pressed into a pellet, and heated at $1200^{\circ}$C for additional 24 hours. After this procedure, a dark-black material was obtained in the form of a 10 mm diameter disk. 

High-resolution XRD data were collected at room $T$ with Cu $K_{\alpha}$ radiation operating at 40 kV and 40 mA. The XRD data was carried over the angular range $10\leq\theta\leq100^{\circ}$, with a 2${\theta}$ step size of 0.01$^{\circ}$. Rietveld refinement was performed with GSAS software, and its graphical interface program \cite{GSAS}. $^{57}$Fe-M\"{o}ssbauer measurement was performed in transmission geometry at room $T$, with a $^{57}$Co:Rh source moving in a sinusoidal motion. Experimental data were fitted with the Normos program. The magnetization ($M$) as a function of $T$ [$M(T)$] and $M$ as a function of $H$ [$M(H)$] measurements were carried out in both zero field cooled (ZFC) and field cooled (FC) modes using a Quantum Design PPMS-VSM magnetometer. The heat capacity data, as well as the ac $M$ as a function of $T$ curves, were also carried in the PPMS, these last being performed in the ZFC mode using the VSM head.

\section{RESULTS AND DISCUSSION}

\begin{figure}
\begin{center}
\includegraphics[width=0.7 \textwidth]{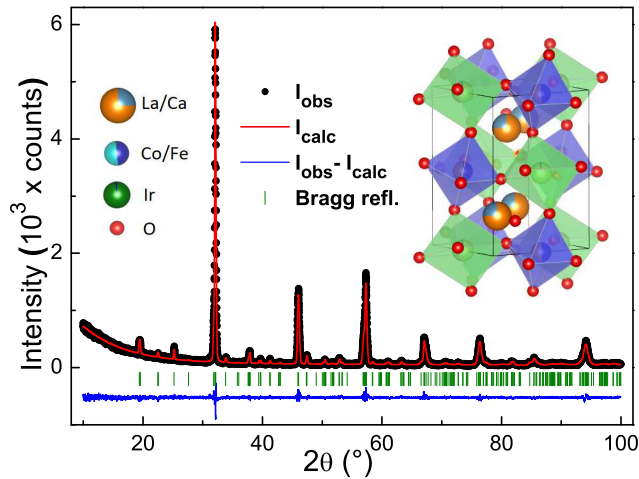}
\end{center}
\caption{Rietveld refinement fitting of LCCFIO. The vertical lines represent the Bragg reflections for the $P2_{1}/n$ space group. Inset shows the crystal structure, in which IrO$_6$ and (Co/Fe)O$_6$ are
drawn as blue and green octahedra, respectively.}
\label{Fig_XRD}
\end{figure}

The room-$T$ XRD pattern of LCCFIO is shown in Fig. \ref{Fig_XRD}. As expected, it is a single-phase DP belonging to monoclinic $P2_{1}/n$ space group, since both La$_{2-x}$Ca$_x$CoIrO$_6$ and La$_{2-x}$Ca$_x$FeIrO$_6$ series form in this same space group \cite{JAP2008,JSSC}. The Rietveld refinement indicates $\sim$10\% of antisite disorder (ASD) at Co/Fe and Ir sites. The similar scattering factors for Co and Fe in Cu $K_{\alpha}$ radiation prevent precise information concerning these ions' individual contributions to the ASD. The main parameters obtained from the refinement are displayed in Table \ref{T1}. 

The lattice parameters here observed are fairly close to those reported for LCCIO, for which a detailed investigation of the oxidation states by means of X-ray absorption spectroscopy (XAS), X-ray magnetic circular dichroism (XMCD) and Raman spectroscopy revealed mixed valence for Co, with $\sim$70\% and 30\% of Co$^{2+}$ and Co$^{3+}$, respectively \cite{PRB2020}. In case of the LCCFIO sample here investigated, M\"{o}ssbauer indicated trivalent state for Fe, while Co and Ir are more likely in bivalent and tetravalent states, respectively, as will be discussed below. This means that the introduction of Fe$^{3+}$ at Co site leads to a slight decrease of the average B-site ionic radius \cite{Shannon}, which may be responsible for the subtle decrease of the unit cell volume (V) of LCCFIO (245.95 \AA$^{3}$) in comparison to LCCIO (247.09 \AA$^{3}$) \cite{PRB2020}. It is established for DP compounds that an increase in the crystal distortion accompanies the volume shrinkage, generally ascribed to tilts in the oxygen octahedra \cite{Sami}, explaining the slightly smaller (Co/Fe)—O—Ir average bond angle of LCCIO with respect to LCCFIO. Such structural and electronic changes will directly impact the systems' magnetic properties.

\begin{table}
\centering
\caption{Main parameters obtained from the XRD, M\"{o}ssbauer, $M(T)$ and $M(H)$ measurements.\label{T1}}
\begin{tabular}{ccccc}
\hline
\hline
$a$ (\AA)    & $b$ (\AA)   & $c$ (\AA)    & $\beta$ ($^{\circ}$)  & V (\AA$^{3}$) \\
5.5587(1)  & 5.6178(2)  & 7.8760(2) & 90.0(1) & 245.95(2) \\
$<$Co/Fe--O--Ir$>$ ($^{\circ}$)  & ASD (\%) & $R_{wp}$ & $\chi^{2}$ \\
148.4(1) & 9.8(2) & 9.7 & 1.6  \\
\hline
   & $\delta$ (mm/s)  & $\Delta$E$_Q$ (mm/s) & $\Gamma$ (mm/s) & A (\%)  \\
$Doublet$: & 0.31  & 0.46  & 0.46  & 100 \\
\hline
$T_{C}$ (K) & $T^{*}$ (K) & $\theta_{CW}$ (K) & $\mu_{eff}$ ($\mu_B$/f.u.) \\
112  & 85  & -118  & 5.3  \\
$M(90 kOe)$ ($\mu_B$/f.u.)  & $M_s$ ($\mu_B$/f.u.) & $M_r$ ($\mu_B$/f.u.) & $H_C$ (kOe) \\
0.56 & 0.32 & 0.30 & 31.7  \\
\hline
\hline
\label{tab1}
\end{tabular}
\end{table}

\begin{figure}
\begin{center}
\includegraphics[width=0.8 \textwidth]{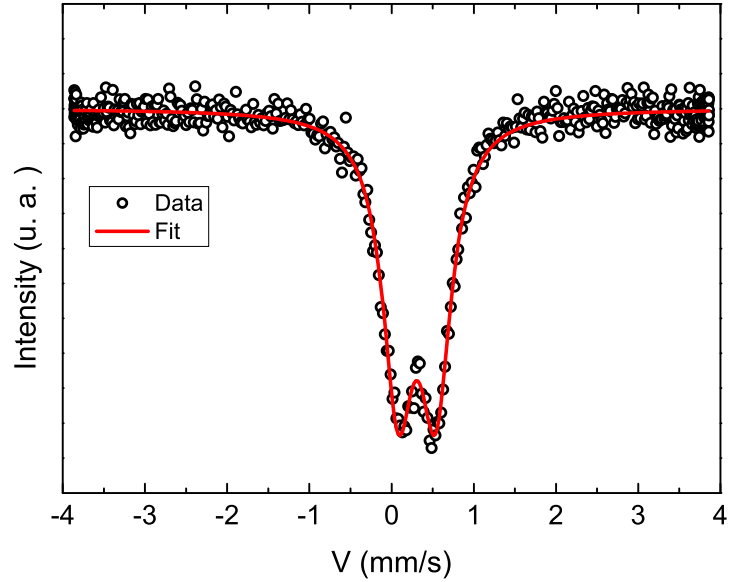}
\end{center}
\caption{${}^{57}$Fe M\"{o}ssbauer spectrum at room temperature showing only one Fe$^{3+}$ site from LCCFIO.}
\label{Fig_Moss}
\end{figure}

$^{57}$Fe M\"{o}ssbauer spectroscopy measurements in transmission geometry were performed on the LCCFIO compound. Fig. \ref{Fig_Moss} shows the M\"{o}ssbauer spectrum for this perovskite at room $T$. The model with only one paramagnetic (PM) subspectrum, doublet, showed the best least-squares fit. This analysis indicates that the compound is in its PM state at this temperature and that the Fe ions prefer only one non-equivalent crystallographic site of the structure. From  M\"{o}ssbauer fit, it was also possible to obtain the hyperfine parameters of isomeric shift $\delta$ (mm/s), quadrupole splitting $\Delta$E$_Q$ (mm/s), line width $\Gamma$ (mm/s) and spectral relative area A (\%). These parameters are displayed in Table \ref{T1}, where the isomer shift value is given in relation to $\alpha$-Fe \cite{greenwood}. The quadrupole splitting value originates from the electric field gradient formed by the atoms surrounding the M\"{o}ssbauer probe ($^{57}$Fe) and may be attributed to an octahedra distortion. A perfect symmetrical octahedron leads to zero quadrupole value. The doublet line width value ($\sim0.46$) may indicate a structural disorder in this compound. Furthermore, the values of the hyperfine $\delta$ and $\Delta$E$_Q$ parameters are consistent with the Fe ions entering the structure with a 3+ oxidation state \cite{greenwood}.

Attempt to perform M\"{o}ssbauer spectroscopy at low temperature in our LCCFIO sample was not successful. 
The spectrum (not shown) taken at 3 K, in a Montana cryofree cryostat for 15 days, is poorly resolved displaying high background. This is probably due to the excitation of X-rays of Iridium near to 14.4 keV M\"{o}ssbauer gamma ray. However, it clearly shows a magnetic sextet with magnetic hyperfine field (Bhf) of 51.4 T and isomer shift (IS) typical of Fe$^{3+}$, as expected. To make any measurements in the magnetic phase in such kind of material one would probably need to use iron enriched in the ${}^{57}$Fe isotope that has natural abundance of 2.8 \%.

\begin{figure}
\begin{center}
\includegraphics[width=0.7 \textwidth]{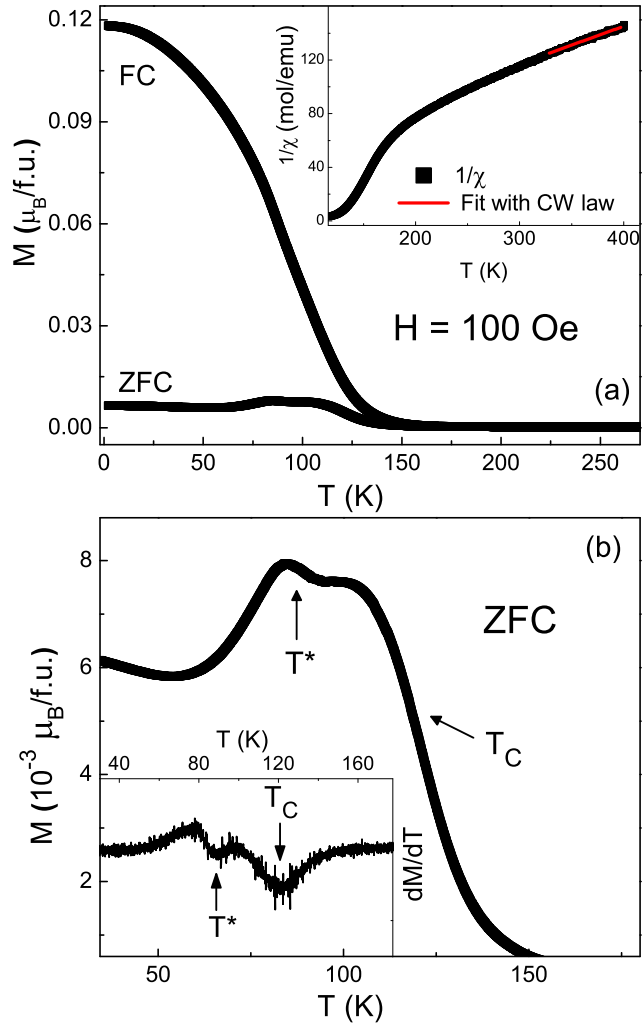}
\end{center}
\caption{(a) ZFC and FC $M(T)$ curves carried at $H$ = 100 Oe. The inset shows the inverse of magnetic susceptibility and the fit of the PM region with the CW law. (b) Magnified view of the ZFC curve around the ordering $T$. The inset shows its first derivative.}
\label{Fig_MxT}
\end{figure}

Fig. \ref{Fig_MxT}(a) displays the ZFC-FC $M(T)$ curves carried at $H$ = 100 Oe, showing a FM-like behavior. However, the small low-$T$ magnetization value rules out the possibility of a fully long ranged FM coupling between Fe, Co and Ir. From the fit of the PM region with the Curie-Weiss (CW) law [see inset of Fig. \ref{Fig_MxT}(a)], we obtained $\theta_{CW}$ = -118 K, fairly close to the magnetic ordering $T$. The negative sign indicates that AFM coupling is dominant. This, together with the distinct magnetic moments expected for the three different TM ions, suggest FIM behavior.

From the fit with CW law we have also obtained the effective magnetic moment, $\mu_{eff}$ = 5.3 $\mu_B$/f.u., which is much larger than the values found for the whole La$_{2-x}$Ca$_x$FeIrO$_6$ series ($\sim$ 4 $\mu_B$/f.u.) \cite{JAP2008} and slightly smaller than that found for La$_{1.5}$Ca$_{0.5}$CoIrO$_6$ (5.4 $\mu_B$/f.u.) \cite{PRB2020}. For the FeIr-based series, M\"{o}ssbauer spectroscopy has shown that Fe remains trivalent for all investigated samples \cite{JSSC}, as for the LCCFIO sample here investigated. On the other hand, for La$_{2-x}$Ca$_x$CoIrO$_6$ a thorough investigation using XAS, XMCD and Raman spectroscopy indicated changes in both Co and Ir oxidation states upon Ca$^{2+}$ to La$^{3+}$ partial substitution \cite{PRB2020}. In order to figure out the electronic configurations of Co and Ir in LCCFIO, one can appeal to the following equation commonly used to compute the theoretical magnetic moment of systems consisting of two or more different magnetic ions \cite{JSSC,Niebieskikwiat}
\begin{equation}
\mu = \sqrt{{\mu_1}^2 + {\mu_2}^2 + {\mu_3}^2 +...}\;. \label{Eq1}
\end{equation}
To ensure charge balance, the TM ions must have a total oxidation state of +6.5 in LCCFIO. Since the M\"{o}ssbauer indicates Fe$^{3+}$ state, we start assuming that Fe$^{3+}$ replaced Co$^{2+}$ in LCCIO, resulting in the La$_{1.5}$Ca$_{0.5}$Co$^{3+}_{0.5}$Fe$^{3+}_{0.5}$(Ir$^{3+}_{0.5}$Ir$^{4+}_{0.5}$)O$_6$ formula. Using the standard magnetic moments for Fe$^{3+}$ ($\mu_{Fe^{3+}}$ = 5.9 $\mu_B$) and HS Co$^{3+}$ ($\mu_{HS Co^{3+}}$ = 5.4 $\mu_B$) \cite{Ashcroft}, and assuming for Ir$^{4+}$ the $\mu_{Ir^{4+}}$ = 1.7 $\mu_B$ value of $J_{1/2}$ state for simplicity ($\mu_{Ir^{3+}}$ = 0), we obtain $\mu$ = 5.8 $\mu_B$/f.u. from Eq. \ref{Eq1}, rather above the experimental value. Since the delicate balance between the crystal field splitting and the intra-orbital Coulomb repulsion in Co$^{3+}$ at octahedral coordination makes its spin state very sensitive to any perturbation \cite{APL2020,Raveau}, we have also checked the above formula with Co$^{3+}$ in low spin (LS) configuration (S = 0), which resulted in $\mu$ = 4.3 $\mu_B$/f.u., far bellow the experimental result. Such discrepancies were expected since Ir$^{3+}$ is rarely observed in octahedral coordination. If we now assume that Fe$^{3+}$ replaces Co$^{3+}$ in LCCIO, the charge neutrality imposes the usual Ir$^{4+}$ configuration, resulting in La$_{1.5}$Ca$_{0.5}$Co$^{2+}_{0.5}$Fe$^{3+}_{0.5}$Ir$^{4+}$O$_6$. Using the standard HS Co$^{2+}$ moment ($\mu_{Co^{2+}}$ = 4.8 $\mu_B$ \cite{Ashcroft}) we get $\mu$ = 5.6 $\mu_B$/f.u., still larger but now closer to the experiment, thus indicating that these are the most likely valence states of the TM ions for LCCFIO. The discrepancy to the experimental result may be related to some overestimation of the Co$^{2+}$ orbital contribution, as well as the naive $\mu_{Ir^{4+}}$ = 1.7 $\mu_B$ assumption for the Ir$^{4+}$ $J_{1/2}$ state. Nevertheless, this is not a unique plausible scenario. A detailed investigation employing XAS, XMCD, and neutron powder diffraction (NPD) is mandatory to determine each ions’ valence and magnetic moment unambiguously.

An accurate inspection of the ZFC curve, Fig. \ref{Fig_MxT}(b), reveals the presence of two anomalies at $T_C$ = 112 K and $T^{*}$ = 85 K, and distinct scenarios can be drawn to explain it. The second anomaly could be related to another magnetic transition apart from $T_C$, in resemblance to the behavior found for other Co- and Fe-based DPs containing three or more magnetic ions \cite{Dass,Zhang,Murthy}. In our case, for instance, the anomalies could be associated respectively to Co$^{2+}$—O—Ir$^{4+}$ and Fe$^{3+}$—O—Ir$^{4+}$ couplings, both predicted by the Goodenough-Kanamori-Anderson rules to be of AFM type. On the other hand, they could be related to the distinct magnetic orderings of roughly independent Co/Fe and Ir AFM sublattices, as recently suggested by DFT calculation in La$_2$CoIrO$_6$ \cite{Ganguly} and corroborated by XMCD and magnetization results in La$_{2-x}$Ca$_x$CoIrO$_6$ \cite{PRB2020}. Another very plausible scenario is the one in which there is only one magnetic transition and the second anomaly would be related to a crossover, \textit{i.e.} some spin reorientation or structural change not necessarily associated with an order parameter. 

\begin{figure}
\begin{center}
\includegraphics[width=0.7 \textwidth]{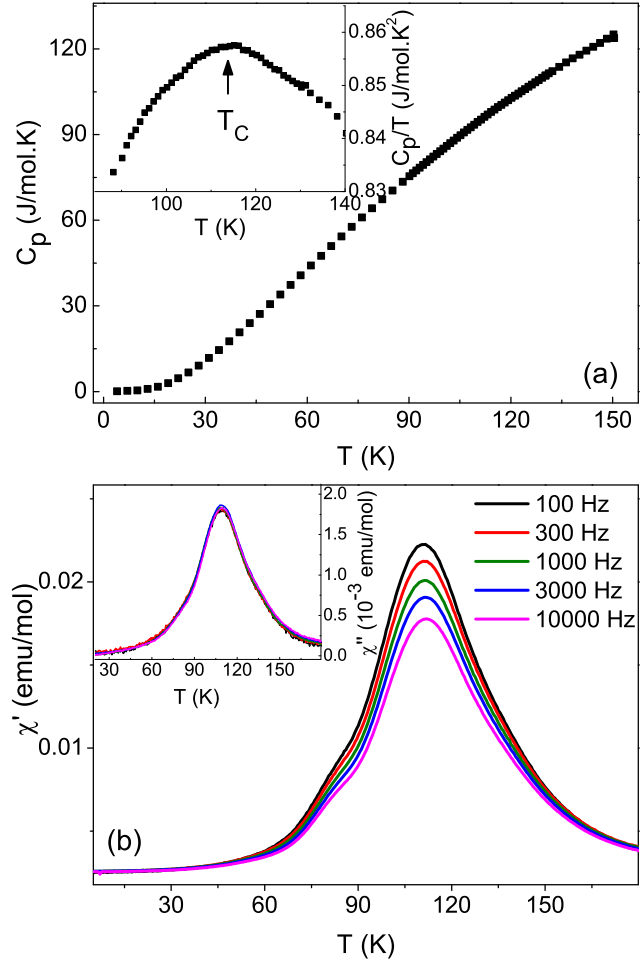}
\end{center}
\caption{(b) Specific heat ($C_p$) as a function of $T$. Inset shows a magnified view of $C_p/T$ near $T_C$. (b) Real part of the ac magnetic susceptibility ($\chi'$) as a function of $T$, measured with $H_{ac}$ = 10 Oe at five different frequencies. Inset shows the imaginary component ($\chi"$).}
\label{Fig_chiAC}
\end{figure}

To unravel this issue, we performed specific heat ($C_p$) measurement as a function of temperature, Fig. \ref{Fig_chiAC}(a). Interestingly, there is no clear transition in the unaided eye $C_p$ curve depicted in the main panel of the figure. However, a magnified view of the $C_p/T$ curve, shown at the inset, indicates a very subtle and broad hump only around $T_C$, suggestive of a disordered system for which competing magnetic interactions and frustration lead to short-range correlations, in a way that the magnonic contribution is masked by the phononic one which becomes relevant at higher temperatures. Similar behavior is commonly observed in disordered DPs \cite{Aczel,Sharma,JSSC2014}. The absence of anomaly at $T^{*}$ indicates that this is not related to any conventional magnetic or structural transition. This is in agreement with previous studies of Ir-based DPs as La$_{2-x}$Sr$_x$CoIrO$_6$ \cite{Narayanan,Lee} and La$_2$ZnIrO$_6$ \cite{Guo}, for which NPD indicate only one magnetic ordering resulting in NCM with two interpenetrating AFM superstructures for Co/Fe and Ir sites, the weak magnetization observed in these compounds coming from spin canting. 

In order to get further insight into the magnetic ordering of LCCFIO, we measured ac $M$ as a function of $T$ with a driving field of 10 Oe and five frequencies ($f$) in the range 100-10000 Hz. Fig. \ref{Fig_chiAC}(b) shows the real ($\chi'$) and imaginary ($\chi"$) parts of the ac magnetic susceptibility  curves, where a peak associated to $T_C$ is clearly observed but at $T^{*}$ there is again only a subtle kink in $\chi'$, which is unnoticed in the $\chi"$ curves. This agrees with the $C_p$ results, further indicating that this anomaly observed in the magnetization data is not related to a second magnetic ordering. However, one can not completely exclude such possibility since its proximity with the broad and intense peak associated with $T_C$ may be preventing the observation of a subtle second transition in the ac susceptibility and $C_p$ curves. It can also be noticed in the ac magnetization measurements that the magnitude of the peaks change with $f$ for both $\chi'$ and $\chi"$ curves. However, there is no systematic shift of their position in $T$, ruling out the possibility of some of the peaks being associated with glassy magnetic behavior. 

Although the resemblance between LCCFIO and the aforementioned La$_{2-x}$Sr$_x$CoIrO$_6$ and La$_2$ZnIrO$_6$ compounds suggests that the most plausible scenario is the one with two AFM sublattices at Co/Fe and Ir sites \cite{Narayanan,Lee,Guo}, one must not take precipitate conclusions since a thorough XMCD study of several distinct combinations of Ir and 3$d$ TM ions in DP compounds showed that the magnetic ground state of these Ir-based DPs is susceptible to the structural and electronic environment \cite{Laguna}. Our magnetization data are not enough to unambiguously determine whether the FIM behavior observed in LCCFIO is due to the AFM coupling between the Co/Fe and Ir FM sublattices or NCM from two interpenetration AFM sublattices for Co/Fe and Ir. For this last scenario, the FIM behavior would result from uncompensated Co--Fe AFM coupling or spin canting. Specific measurements such as temperature-dependent XRD, XMCD, and NPD would be necessary to unravel the magnetic structure of LCCFIO.

\begin{figure}
\begin{center}
\includegraphics[width=0.7 \textwidth]{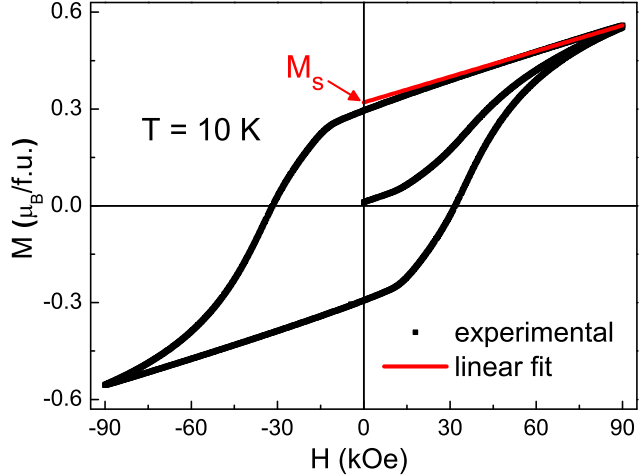}
\end{center}
\caption{$M(H)$ loop at 10 K. The solid red line is an extrapolated linear fit of the high $H$ data ($>$80 kOe).}
\label{Fig_MxH}
\end{figure}

The $M(H)$ curve can give us further insight into the magnetic ground state of LCCFIO. Fig. \ref{Fig_MxH} displays the hysteresis loop carried out at $T$ = 10 K. It is a closed-loop, symmetric with respect to both $M$ and $H$ axes, with an FM-like shape. However, the lack of saturation even at $H$ = 90 kOe further indicates AFM coupling between the TM ions, resulting in FIM behavior. The main results extracted from the curve are displayed in Table \ref{T1}. 

The $M_s$ = 0.32 $\mu_B$/f.u., obtained from the extrapolation of a linear fit of the high $H$ data ($>$ 80 kOe), as well as the $M$ = 0.56 $\mu_B$/f.u. at $H$ = 90 kOe (see Table \ref{T1}), are far below the value expected for a simple linear AFM coupling between two FM sublattices in Ir and Co/Fe sites, even considering the $\sim$10\% of ASD. As aforementioned, NPD studies of La$_{2-x}$Sr$_x$CoIrO$_6$ indicated that Co and Ir form two AFM sublattices \cite{Narayanan,Lee}. Assuming a similar scenario here, \textit{i.e.} an AFM sublattice for Ir and another for Fe/Co, the Ir$^{4+}$ magnetic moments (here considered simply as $J$ = 1/2) would cancel. However, the AFM coupling between Fe$^{3+}$ (S = 5/2) and Co$^{2+}$ (S = 3/2) would not compensate, resulting in FIM behavior. Taking into account that we have 0.5 of Fe$^{3+}$ and 0.5 of Co$^{2+}$ per formula unit, the difference between the moments yields $M$ = 1.0 $\mu_B$/f.u., somewhat larger than $M_s$ and $M(90 kOe)$. Applying the usual formula to compute the decrease in $M$ due to the ASD \cite{Sami,Feng}
\begin{equation}
M = M_{exp}\times(1-2ASD), \label{Eq2}
\end{equation}
where $M_{exp}$ is the theoretical SO moment, one has $M$ = 0.8 $\mu_B$/f.u., now closer but still larger than the experiment. Such discrepancy to the value expected for a linear Co--Fe AFM coupling gives evidence toward the spin canting scenario, since the small net magnetization values depicted in Table \ref{T1} are of the same order of magnitude as those reported for spin canted Ir-based DPs \cite{Narayanan,Guo}. However, the possibility of NCM with two distinct AFM superlattices for the Fe/Co and Ir sites can not be completely excluded since other ingredients besides ASD may contribute to the further decrease of the magnetization, as for instance  antiphase boundaries, oxygen vacancies and defects that could lead to local changes in the TM ions valences, as the afore discussed possible presence of some small amount of Co$^{3+}$. Again, further investigation is necessary to determine whether the FIM-like behavior of LCCFIO results from uncompensated Co--Fe AFM coupling, from spin canting or even from a simple AFM coupling between Co/Fe and Ir FM sublattices.

Another interesting result obtained from the $M(H)$ curve is the large $H_C$ $\simeq$ 32 kOe, comparable to those of other DPs of interest for hard magnets \cite{Feng,Jansen}, and much larger than that reported for LCCIO \cite{PRB2020}. Previous studies of chemical pressure on La$_{1.5}A_{0.5}$CoMnO$_6$ ($A$ = Ba, Ca, Sr) report the increase of $H_C$ with the decrease of the average $A$-site ionic radius, ascribed to the enhancement of the orbital contribution to the magnetic moments caused by the increased lattice distortion \cite{PRB2019}. Conversely, a recent investigation of hydrostatic pressure on $A_2$FeReO$_6$ ($A$ = Ba, Ca) has shown a dramatic increase of $H_C$ attributed to pressure-induced changes in the crystal field rather than in the orbital moment \cite{Haskel}. For $A_2$NiIrO$_6$ ($A$ = La-Lu), a systematic increase of $H_C$ and $T_C$ was observed with decreasing the unit cell volume, ascribed to the enhanced Ni $e_g$-Ir $t_{2g}$ orbital hybridization due to lattice distortion \cite{Feng}. In our case, the great increase in the $H_C$ of LCCFIO with respect to that of LCCIO is certainly not related to the orbital moment of 3$d^5$ Fe$^{3+}$ inserted in the system, presumably negligible, but probably to its spin moment and also to the larger structural distortion of the former compound in comparison to the later one, which signifies stronger $t_{2g}$-$t_{2g}$ orbital hybridization. 

\section{CONCLUSIONS}

In summary, the polycrystalline LCCFIO sample here investigated is a single-phase DP formed in the monoclinic $P2_{1}/n$ space group with the presence of $\sim$10$\%$ of ASD at Co/Fe and Ir sites. M\"{o}ssbauer spectroscopy analysis revealed that Fe is in a trivalent oxidation state, while the magnetization results indicate mixed-valence for Co and Ir. The magnetization data also revealed a FIM behavior below 112 K, possibly due to the presence of two AFM superstructures for Co/Fe and Ir ions, with the weak net magnetization coming from the uncompensated coupling in the Co/Fe lattice or from spin canting. The significant increase in $H_C$ for LCCFIO with respect to that of LCCIO was discussed in terms of the structural distortions and electronic changes induced by the partial substitution of Co by Fe. Such large $H_C$ is comparable to other DPs of interest for hard magnets, being worth further investigation.

\section{ACKNOWLEDGMENTS}

This work was supported by Conselho Nacional de Desenvolvimento Cient\'{i}fico e Tecnol\'{o}gico (CNPq) [No. 425936/2016-3], Coordena\c{c}\~{a}o de Aperfei\c{c}oamento de Pessoal de N\'{i}vel Superior (CAPES)and Funda\c{c}\~{a}o de Amparo \`{a} Pesquisa do Estado de Goi\'{a}s (FAPEG). E.B.S. aknowledges several grants from Funda\c{c}\~{a}o Carlos Chagas Filho de Amparo \`{a} Pesquisa do Estado do Rio de Janeiro (FAPERJ), including Professor Emeritus fellowship.

\end{document}